\documentclass[fleqn,usenatbib]{mnras}

\usepackage{newtxtext,newtxmath}

\usepackage[T1]{fontenc}

\DeclareRobustCommand{\VAN}[3]{#2}
\let\VANthebibliography\thebibliography
\def\thebibliography{\DeclareRobustCommand{\VAN}[3]{##3}\VANthebibliography}

\usepackage{xspace}
\usepackage{enumitem}
\usepackage{todonotes}
\usepackage{orcidlink}
\usepackage{gensymb}

\graphicspath{{./figs/}}

\newcommand{\src}{{\it GOTO 0650}}

\title[Revealing the Hidden Spin of the WZ Sge Star {\it GOTO 0650}]{Bridging the Gap: {\it OPTICAM} Reveals the Hidden Spin of the WZ Sge Star {\it GOTO 065054.49+593624.51}}

\author[N. Castro Segura et al.]{N.~Castro~Segura$^{\orcidlink{0000-0002-5870-0443}}$\,$^{1,2}$\thanks{E-mail: N.Castro-Segura@warwick.ac.uk},
Z.~A.~Irving$^{\orcidlink{0009-0006-0951-3429}}$\,$^{2}$, 
F.~M.~Vincentelli$^{\orcidlink{0000-0002-1481-1870}}$\,$^{2}$,
D.~Altamirano$^{\orcidlink{0000-0002-3422-0074}}$\,$^{2}$,
Y.~Tampo$^{\orcidlink{0000-0003-0196-3936}}$\,$^{3,4}$,
C.~Knigge$^{\orcidlink{0000-0002-1116-2553}}$\,$^{2}$, \newauthor
I.~Pelisoli$^{\orcidlink{0000-0003-4615-6556}}$\,$^{1}$,
D.~L.~Coppejans$^{\orcidlink{0000-0001-5126-6237}}$\,$^{1}$
N.~Rawat$^{3}$, 
A.~Castro$^{\orcidlink{0000-0002-7832-5337}}$\,$^{5,2}$, 
A.~Sahu$^{\orcidlink{0009-0007-6825-3230}}$\,$^{1}$,
J.~V.~Hern\'andez Santisteban$^{\orcidlink{0000-0002-6733-5556}}$\,$^{6}$,\newauthor
M.~Kimura$^{8}$,
M. Veresvarska$^{\orcidlink{0000-0002-0146-3096}}$\,$^{7}$,
R.~Michel$^{\orcidlink{0000-0003-1263-808X}}$\,$^{5}$, 
S. Scaringi$^{\orcidlink{0000-0001-5387-7189}}$\,$^{7,10}$ 
 \&
M.~R.~Najera$^{\orcidlink{0000-0003-3283-0407
}}$\,$^{5}$.
\newauthor
\emph{\normalsize Affiliations are listed at the end of the paper}
}

\date{Accepted XXX. Received YYY; in original form ZZZ}

\pubyear{2025}

\begin{document}
\label{firstpage}
\pagerange{\pageref{firstpage}--\pageref{lastpage}}
\maketitle

\begin{abstract}
WZ Sge stars are highly evolved accreting white dwarf systems (AWDs) exhibiting remarkably large amplitude outbursts (a.k.a. super-outbursts), typically followed by short rebrightenings/echo outbursts. 
These systems have some of the lowest mass transfer rates among AWDs, making even low magnetic fields dynamically important. 
Such magnetic fields are often invoked to explain the phenomenology observed in these systems, such as their X-ray luminosity and long periods of quiescence (30+ years).
However, the detection of these is very elusive given the quenching of the accretion columns during outburst and the low luminosity of these systems during quiescence. 
Here we present high-cadence multi-band observations with {\it OPTICAM} of the recent outburst of the recently discovered
WZ Sge star GOTO065054.49+593624.51, during the end of the main outburst and the dip in-between rebrightenings, covering 2 orders of magnitude in brightness. 
Our observations reveal the presence of a statistically significant signal with $P_{\omega}\simeq148$ seconds in the bluer ($g$) band which is detected only during the dip between the main outburst and the rebrigthenings. We interpret this signal as the spin period of the AWD.
If confirmed, GOTO 0650 would bridge the gap between intermediate- and fast-rotating intermediate polars (IPs) below the period gap. 
\end{abstract}





\section{Introduction} \label{sec: intro}
Cataclysmic variable stars (CVs) are binary systems in which a low-mass secondary star, filling its Roche lobe, transfers mass to a white dwarf (WD) primary. In systems where the magnetic field is dynamically unimportant, the accretion process is mediated by an accretion disc. If the magnetic field of the WD is sufficient to prevent the formation of the disc, this gives rise to the so-called {\it polars}. If, on the other hand, a disc can form but is truncated by the magnetosphere of the central WD, the systems are traditionally classified as {\it intermediate-polars} (IPs), however, recent findings suggest that some IPs may be disc-less \citep[e.g. ][]{Littlefield:2021AJ....162...49L}. Disc-mediated systems with mass transfer rates sufficiently low to prevent the disc from remaining persistently in a fully ionized state are classified as Dwarf novae \citep[DNe; see e.g.][ for a review on CVs]{Warner1995cvs..book.....W}. DNe are characterized by exhibiting eruptions where the system brightens typically 2-5 magnitudes for a relatively short period of time. Such eruptions are thought to be the consequence of a sudden increase in the mass accretion rate onto the compact object due to the accretion disc transitioning from a neutral to a fully ionized state \citep[e.g.,][]{Lasota2016}.

WZ Sge-systems are highly-evolved short orbital period DNe 
which exhibit even larger amplitude outbursts (up to $\sim 8-9$ mag). These so-called super-outbursts are characterized by prolonged outburst durations (approximately 30 days), and the appearance of superhumps — low-amplitude variations near the system's orbital period — shortly after reaching peak brightness \citep[][]{smak93, smak09a, Patterson_SHs_2005PASP..117.1204P, Kato2015PASJ...67..108K,hameury20}. These super-outbursts are typically followed by a series of short-term rebrightenings or echo outbursts which seem to be characteristic of accreting binaries with extreme mass-ratios. 
These rebrightenings cannot be explained by the standard disc instability model \citep[DIM; e.g. ][]{hameury20}. To reconcile observations with theoretical predictions, \citet{Campana:2018A&A...610A..46C} recently proposed an alternative explanation where the luminosity drops during outburst decay signal a temporary transition to a magnetic propeller state. 
This is an appealing framework as it could help to explain the behaviour of accreting systems at different scales, i.e. X-ray binaries, young stellar objects and CVs. 
This framework also aligns well with the proposed spin period of 27.87 seconds of {\it WZ Sge} \citep{Patterson1980,Patterson1998}. However, using HST time-resolved spectroscopy, \citet{Georganti2022MNRAS.511.5385G} showed that {\it WZ Sge} does not exhibit any propeller footprint during the proposed propeller state \citep[c.f.][]{EracleousHorne:1996}. Nonetheless, truncation of the inner disc is required to explain the relatively high X-ray luminosity. Against this background, \cite{hameury20} proposed that the X-ray excess observed in {\it WZ Sge} could (also) be a consequence of evaporation of the inner disc, while the quiescent phases between outbursts are produced by overflows of the mass-transfer stream impacting the inner regions of the disc, similar to what is proposed to explain IW And-systems and some nova-like CVs \citep{Kimura2020PASJ..tmp..156K,CastroSegura:2021MNRAS.501.1951C}. Regardless of whether magnetic fields are responsible for the X-ray luminosities in WZ Sge stars, they are often invoked to explain the very long recurrence timescales in these systems \citep{Kato2015PASJ...67..108K}.

\begin{figure*}
      \centering
      \includegraphics[width=0.95\textwidth]{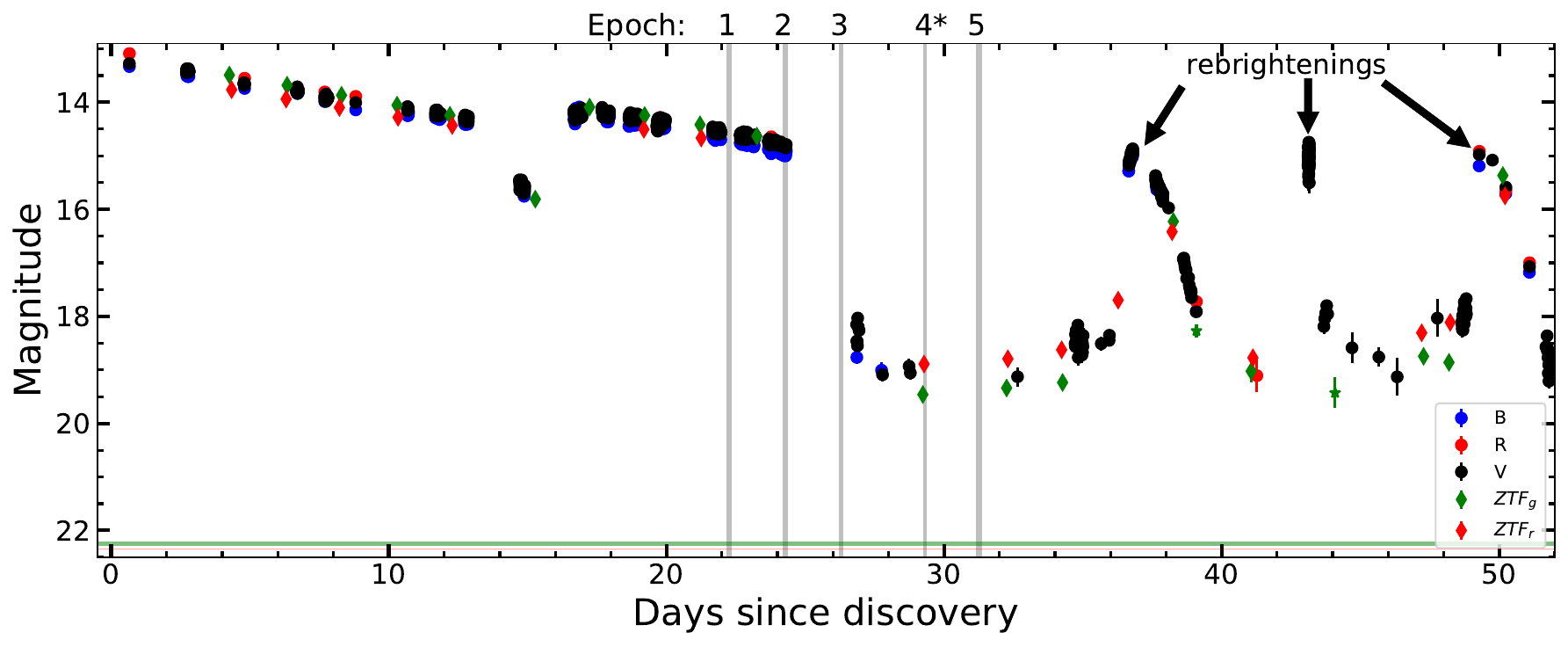}
    \caption{Outburst evolution of \src\ as reported by the AAVSO (circles) and ZTF (diamonds for psf- and stars for forced-photometry). The different colours indicate the magnitude in different bands. The coloured horizontal bands encompass the $\pm1\sigma$ error around the weighted mean of the quiescent photometry obtained with forced photometry on ZTF during 9-25 days before the onset of the superoutburst, the precise values are $g = 22.25\pm 0.03$ mag and $r = 22.351 \pm 0.005$ mag. The arrows highlight the maxima of the post-outburst re-brightening/-flares. The time of the high-cadence observations with {\it OPTICAM} are indicated with the grey shaded regions, the epoch ID for each observation reported in this paper is indicated on the top axis. (*) denotes the epoch with a statistically significant detection of the $148$s signal.}
    \label{fig: outburst evo}
\end{figure*}

In this Letter we report high cadence multi-band photometry of the newly discovered transient {\it GOTO065054.49+593624.51} (hereafter \src) obtained with {\it OPTICAM} \citep{OPTCIAM2019RMxAA,OPTICAM2024}. \src\ was discovered by the Gravitational-wave Optical Transient Observer \citep[GOTO;][]{ GOTO_Steeghs2022MNRAS.511.2405S,GOTO_2024SPIE13094E..1XD} on  Oct. 4 2024 03:36:36 UT \citep{Killestein_discovery2024ATel16842....1K}. The transient reached an L band ($\sim g+r$) magnitude of 13.7 in the discovery images and was associated with a $\sim 22^{nd}$ mag quiescent counterpart implying an outburst amplitude of $\sim 8.5$ mag, typical of WZ Sge stars. The nature of the transient was confirmed by spectroscopic observations and ultraviolet photometry \citep[e.g.][]{Killestein_spectra_2024ATel16858....1K, Bhattacharya2024ATel16866....1B}. 
Using data from the {\it American Association of Variable Star Observers} ({\it AAVSO}) and the {\it Zwicky Transient Facility} \citep[{\it ZTF};][]{ZTF_2019PASP..131a8002B,ZTF2_2019PASP..131f8003B}, we show in Figure \ref{fig: outburst evo} how the outburst evolution and rebrigthenings are consistent with the classification above. 
\citet{Tampo24_vsnet} reported the onset of regular super-humps after a type-E recovery of a dip $\sim15\,{\rm d}$ from the beginning of the outburst \citep[e.g.][]{Kimura2018PASJ...70...47K}, and suggested \src\ is a period bouncer with an orbital period close to the superhump period (${\rm P_{sh}} = 91.05\pm3$ min), again, in line with \src\ being a member of the WZ Sge-type DNe.

\section{Observations and data reduction}\label{sec: obs}


The {\it OPtical TIming CAMera} \citep[{\it OPTICAM;}][]{OPTCIAM2019RMxAA,OPTICAM2024} is a new high-cadence, multi-band camera mounted on the 2.1-meter telescope at the {\it San Pedro M\'artir Observatory (OAN-SPM)}, in M\'exico. {\it OPTICAM} is equipped with three Andor Zyla 4.2-Plus sCMOS cameras with three $2048\times2048$ pixels and a set of SDSS filters ($ugriz$), allowing coverage in the $3200 {\rm AA} -1.1{\rm \mu m}$ range. The  field of view (FoV), is $ \approx 5\times 5~{\rm arcmin^2}$ with a pixel-scale of $\simeq 0.15 ''/{\rm pix}$.

\src\ was observed in the $g,r,i$ bands during the nights of October 26, 28, 30 and November 2$^{nd}$ and 4$^{th}$ of 2024. The sky conditions were photometric for all the nights except for the last one, during which variable high clouds were present; this is reflected in our ability to recover the i band photometry in epoch 5.
The details of the observations can be found in Table \ref{tab: obs}. The data were primarily reduced using version 1.13.0 of the \texttt{photutils} Python package \citep{larry_bradley_2024_12585239}. Cosmic rays were clipped from all images using the L.A.Cosmic algorithm \citep{LACOSMIC2001PASP..113.1420V} as implemented in \texttt{Astro-SCRAPPY} version 1.2.0 \citep{McCully2018}. For each image, two dimensional background images were calculated using the \texttt{SExtractorBackground} estimator from \texttt{photutils} with a \texttt{box\_size} of 64 pixels for $2 \times 2$ binning or 42 pixels for $3 \times 3$ binning. Image segmentation was used to identify sources in the background and subtracted images using the \texttt{SourceFinder} routine from \texttt{photutils}, which combines source detection and deblending. To identify sources, we set the \texttt{npixels} parameter to 32 pixels for $2 \times 2$ binning and 14 pixels for $3 \times 3$ binning. The \texttt{threshold} parameter was set to $5 \times {\rm background~RMS}$, and the background RMS was estimated using \texttt{StdBackgroundRMS}. All other parameters were left to their default values.

We used the optimal photometry algorithm described in \cite{Naylor1998} to perform photometry on the background-subtracted images. We model the PSF of {\it OPTICAM}'s three cameras using two-dimensional Gaussians for each camera. We stacked the images from each epoch for each camera to get the PSF parameters. We then identified sources in the stacked images using \texttt{SourceFinder}, which assigns a semi-major and semi-minor standard deviation to each source. The PSF for each camera in each epoch was then modelled using the median semi-major and semi-minor standard deviations from the corresponding stacked image. We then used differential photometry to correct for atmospheric variability; relative light curves were computed using the summed fluxes of the same (non-variable) reference stars in each night.

\begin{table}
	\centering
	\caption{Observing log. The star in the epoch number denotes statistically significant detection of short-term periodicity (see Sec. \ref{sec: analysis} for details).}
	\label{tab: obs}
	\begin{tabular}{lccccr} 
		\hline
		Epoch & MJD$_{\rm start}$ & filters & exposure time& obs. length& binning\\
		\hline
        \ & day & \ & seconds & hours &
        \\
		\hline
        \hline
        $1$& $ 60609.31 $& $gri$ & $3$ & $4.65$ & $2\times2$ \\
        $2$& $ 60611.33 $& $gri$ & $3$ & $4.54$ & $2\times2$ \\
        $3$& $ 60613.38 $& $gri$ & $5$ & $3.62$ & $2\times2$ \\
        $4^*$& $ 60616.39 $& $gri$ & $15$ & $3.34$ & $3\times3$ \\
        $5$& $ 60618.29 $& $gri$ & $15$ & $5.89$ & $3\times3$ \\
        \hline
	\end{tabular}
\end{table}

\section{Analysis and results}  \label{sec: analysis}

\begin{figure*}
    \centering
    \includegraphics[width=.87\textwidth]{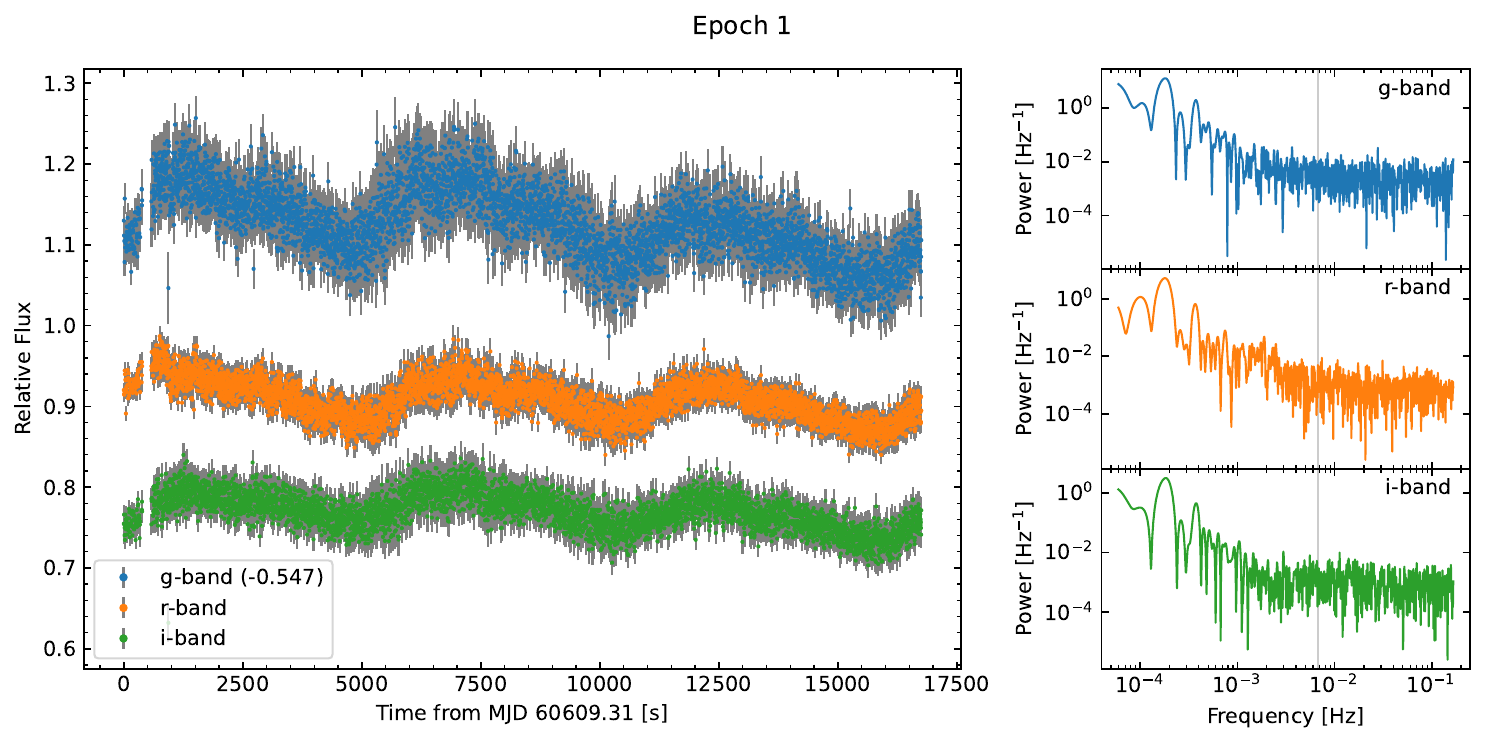}
    \includegraphics[width=.87\textwidth]{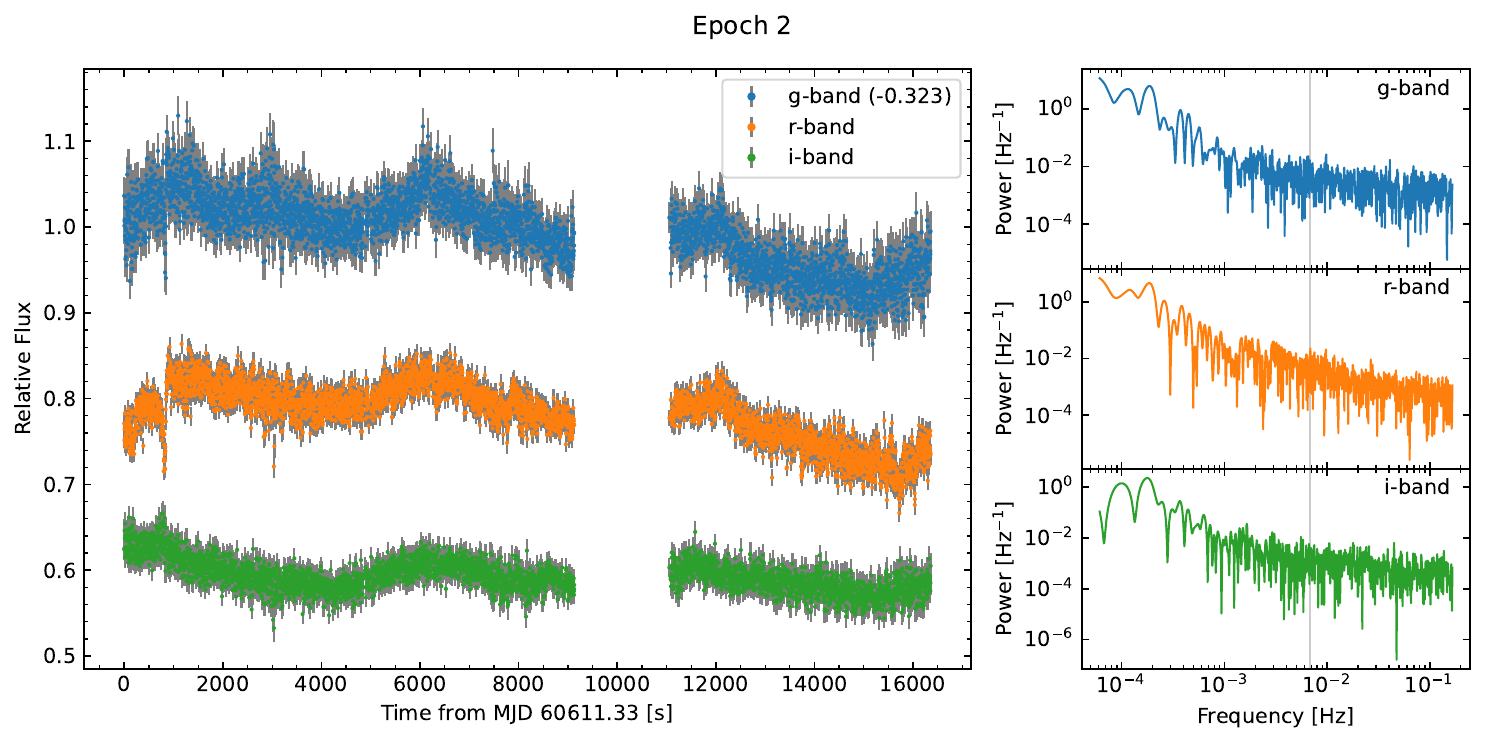}
    \includegraphics[width=.87\textwidth]{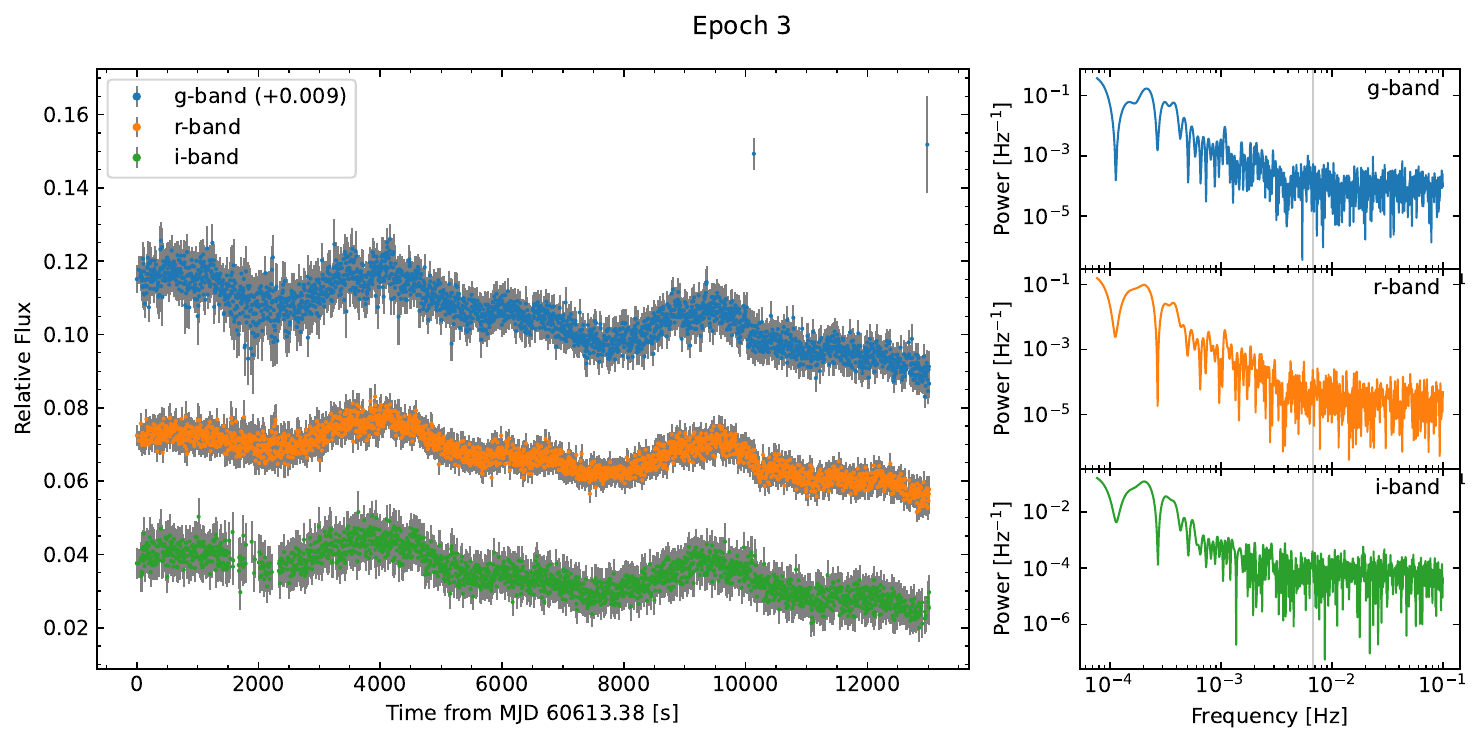}
    \caption{Relative light curves of all three cameras for epochs $1-3$ (left panels), along with their respective periodograms (right panels). The vertical shaded region in the power spectra indicates the frequency at which we detect a statistically significant signal in epoch 4.}
    \label{fig: epoch 1-3}
\end{figure*}

\begin{figure*}
    \centering
    \includegraphics[width=.87\textwidth]{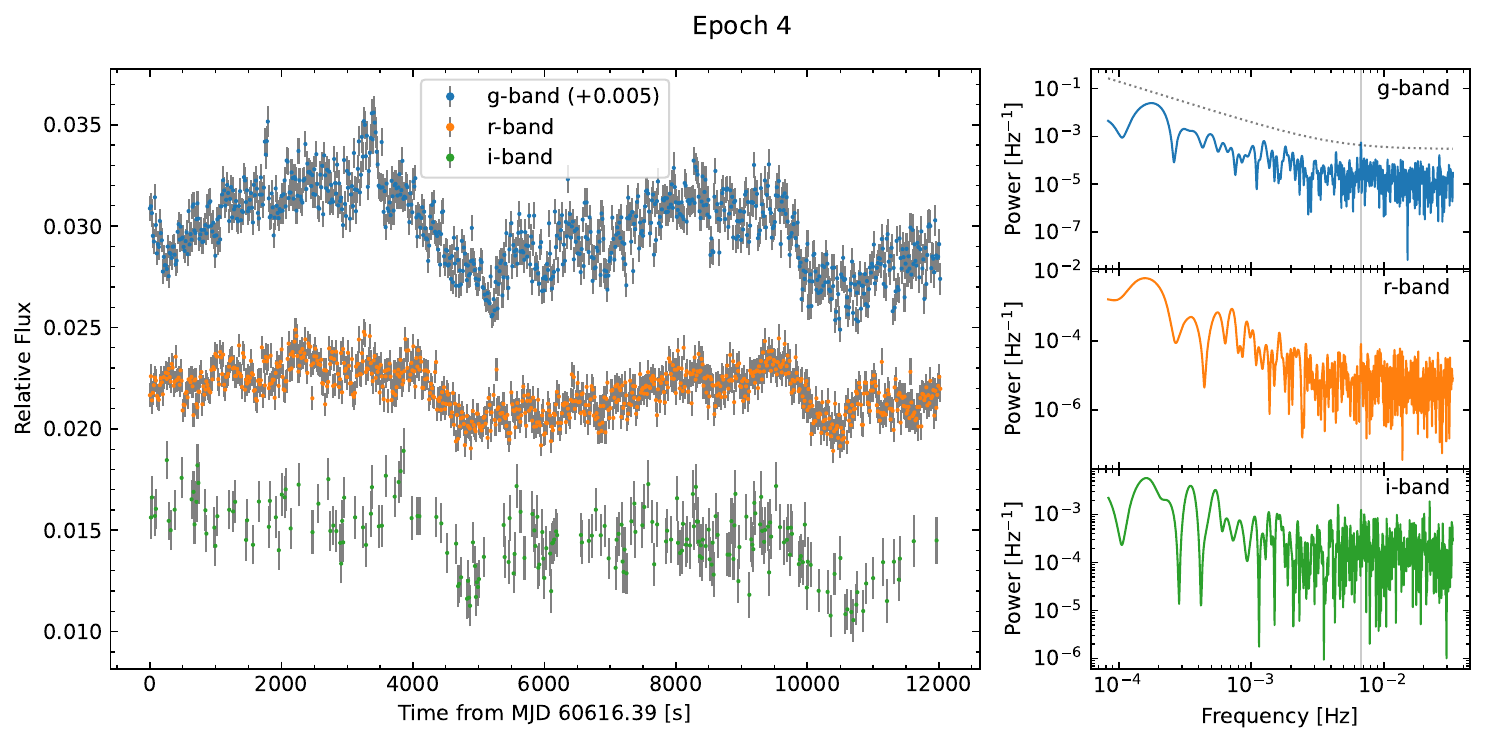}
    \includegraphics[width=.87\textwidth]{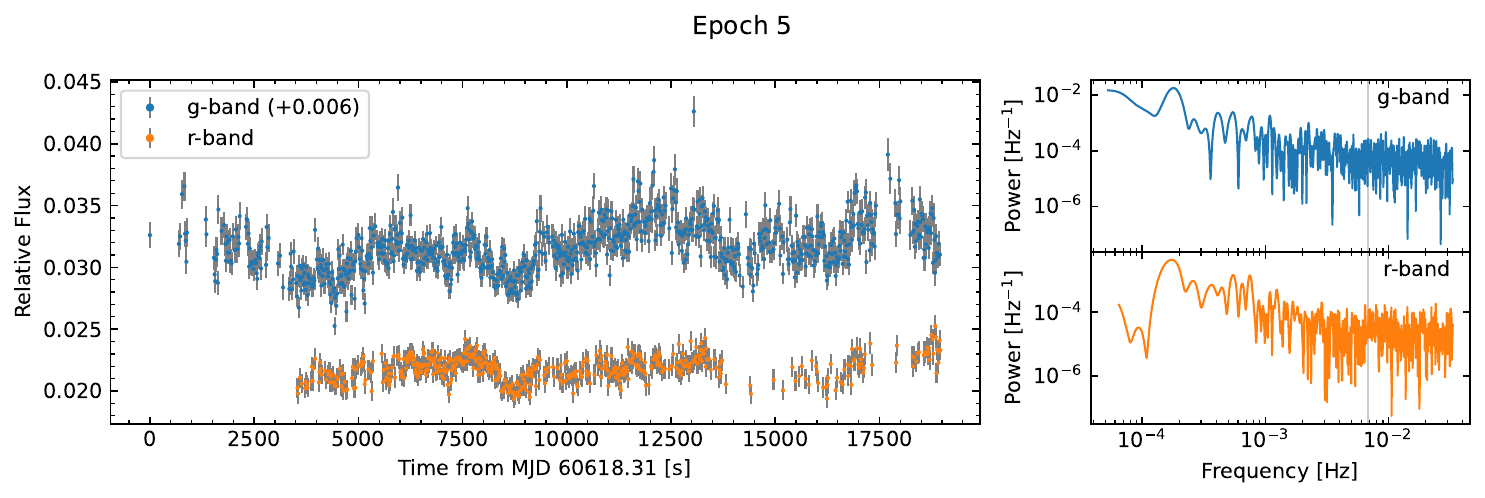}
    \caption{Same as Fig. \ref{fig: epoch 1-3} but for epochs 4 and 5. 
    For the $g$-band periodogram in epoch 4, the 99.99 per cent confidence threshold is shown with a dotted line. The target was not detected in the $i$ band during epoch 5, therefore, only g and r bands are shown.}
    \label{fig: epochs 4 5}
\end{figure*}

The high cadence photometric light curves are presented in the left panels of Figures \ref{fig: epoch 1-3} and \ref{fig: epochs 4 5}, the comparison stars in these figures are always the same (namely Gaia EDR3 1003219393009879424 and 1003225264228638336), so the change in relative flux between epochs is meaningful. As can be seen in Figure \ref{fig: outburst evo}, epochs 1 and 2 were taken towards the end of the superoutburst, epoch 3 was taken during the decline, epochs 4 and 5 occurred during the dip between the main outburst structure and the rebrightenings, when the source was $\sim 4-5$ mags fainter than the first observation but $\sim 3$ mags brighter than the quiescent level. During epochs 1--3, the source was dominated by the regular superhumps reported by \citet{Tampo24_vsnet}, with the dispersion in the light curves reducing as the accretion disc got fainter. In epochs 4 and 5 there is still a slow modulation present with similar frequency as before, as can be seen in Figure \ref{fig: epochs 4 5}, the general trend is similar but the $g$-band exhibits higher amplitude (than at least the $r$-band). In both of these epochs, we can see short-term variability in the $g$-band, suggesting the presence of an additional component contributing to the modulation of the light curve.  

To characterise the short term variability of \src\ we have computed the generalized Lomb-Scargle periodogram \citep[LSP;][]{Lomb1976,Scargle1982,VanderPlas2018ApJS..236...16V} for each light curve as implemented in \texttt{astropy} version 6.1.2. We normalised our LSPs according to:
\begin{equation}
    P = {2 \delta t} P_{\rm LSP},
    \label{eq: LSP normalisation}
\end{equation}
where $\delta t$ is the time resolution of the light curve and $P_{\rm LSP}$ represents the {\it unnormalised} LSP powers. This normalisation is defined such that the integrated periodogram yields the variance of the light curve\footnote{Provided the time series is uniformly sampled and the LSP is evaluated at the Fourier frequencies.} \citep{Vaughan2003}.

To estimate the confidence levels of the proposed signals in our LSPs, we model the continuum analytically following the method described in Appendix A of \cite{Vaughan2005}, which we note is only applicable to the LSP if the time series is uniformly sampled and the LSP is evaluated at the Fourier frequencies \citep{vanderplas2018}. We compute confidence thresholds by scaling our continuum model by $- 2 {\rm ln}(\epsilon / n')$, where $\epsilon$ is the desired false alarm probability and $n'$ is the number of LSP ordinates used to fit the model. The LSP for each of the light curves is shown in the right panel of Figures \ref{fig: epoch 1-3} and \ref{fig: epochs 4 5}. We found a periodicity of $\rm P_{\omega} = 148.2$ s with more than $99.99$ per cent confidence only in the $g$-band of epoch 4, though we also note a non-statistically significant signal at the same frequency in the r band of the same epoch (72.25 per cent confidence).  
In turn, we are not able to recover any signal from the LSP but, despite this, the $g$-band lightcurve from that same epoch exhibits short-term flares qualitatively similar to those observed in epoch 4. We derive the analytical false alarm probability of our epoch 4 $g$-band detection to be $1.16 \times 10^{-6}$; the probability of detecting a signal of this significance in at least one of our 14 LSPs by chance (i.e., when there is no underlying signal) is 0.0016 per cent.

To determine the centre and $1\sigma$ uncertainty of the frequency detected in epoch 4, and the coherence of this modulation, we performed a bootstrap analysis \citep[e.g.][]{Stats_ML_DataMining2014sdmm.book.....I}. We note that bootstrapping can also be used to reject sampling aliases \citep[e.g.,][]{Southworth2006}, which may be present in the $i$-band LSP. From this analysis, we found peak frequencies of $6.73 \pm 0.02$~mHz ($g$-band), $6.73 \pm 0.02$~mHz ($r$-band), and $6.75 \pm 0.02$~mHz (i-band). These frequencies correspond to periods of $148.5 \pm 0.4$~s, $148.7 \pm 0.4$~s, and $148.2 \pm 0.5$~s, respectively. To further test the coherence, we fit a Lorentzian to the Epoch 4 $g$-band periodogram \citep{Belloni2002} on top of our noise model \citep[][top right panel in Figure \ref{fig: epochs 4 5}]{Vaughan2005} and found a Q factor of 3422. The bootstrapped uncertainties and high Q factor are both indicative of a highly coherent signal.

\begin{figure}
    \centering
    \includegraphics[width=\columnwidth]{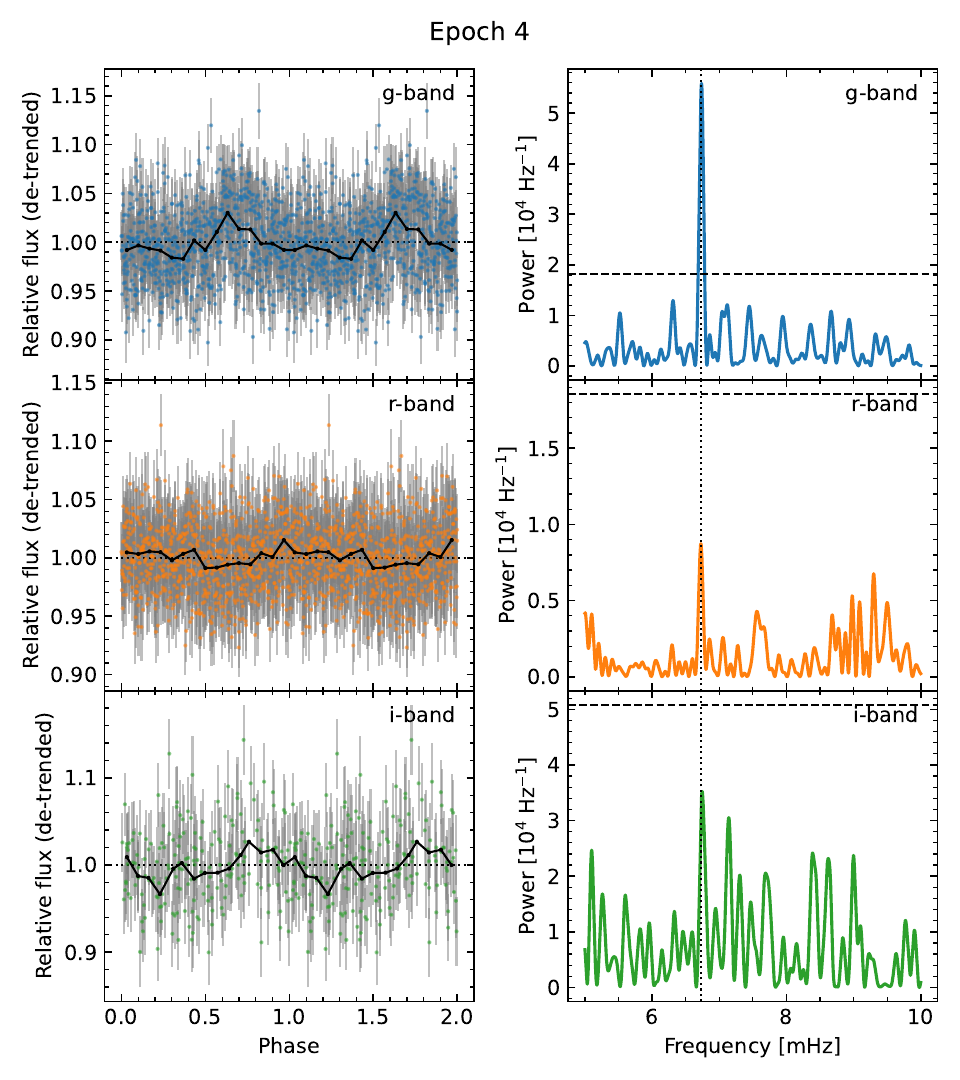}
    \caption{Left: Epoch 4 light curves folded on a period of 148.6 s (corresponding to 6.7 mHz), the black line represents the median values for the phase binned folded light curve, the zero phase is arbitrary but common across bands. Right: Epoch 4 LSPs. In the LSPs, the dotted vertical line corresponds to a frequency of 6.7~mHz and the horizontal line represents the $99.99$ per cent confidence threshold.}
    \label{fig: epoch 4 phase fold}
\end{figure}

To visualise the modulations identified in Epoch 4, we phase folded these light curves on the peak frequencies inferred via our bootstrap analysis; we present these phase folded light curves in Figure \ref{fig: epoch 4 phase fold}, along with localised LSPs. Prominent modulations are only apparent in the g-band, with a peak-to-peak amplitude of roughly 10 per cent (compared to an average $1\sigma$ white noise amplitude of 2.6 per cent). The $3\sigma$ upper limits using the average white noise amplitude for the Epochs 1, 2, 3 and 5 in $g$-band is 4.8, 4.5, 14.5 and 10.5 per cent respectively. For the LSPs, 99.99 per cent confidence thresholds were estimated by simulating 10,000 white-noise light curves for each band. Each simulated light curve had the same time sampling as the observed light curve, and fluxes were generated from the following Gaussian $f_i \sim \mathcal{N}(\Bar{f}, \sigma_{f_i})$, where $\Bar{f}$ represents the mean flux of the observed light curve and $\sigma_{f_i}$ represents the uncertainties on the observed flux values. As can be seen in Figure \ref{fig: epochs 4 5}, the periodograms of Epoch 4 are white-noise-dominated beyond 5~mHz, and so our estimated confidence thresholds are unlikely to be significantly biased in this frequency range.

\begin{figure}
    \centering
    \includegraphics[width=0.9\columnwidth]{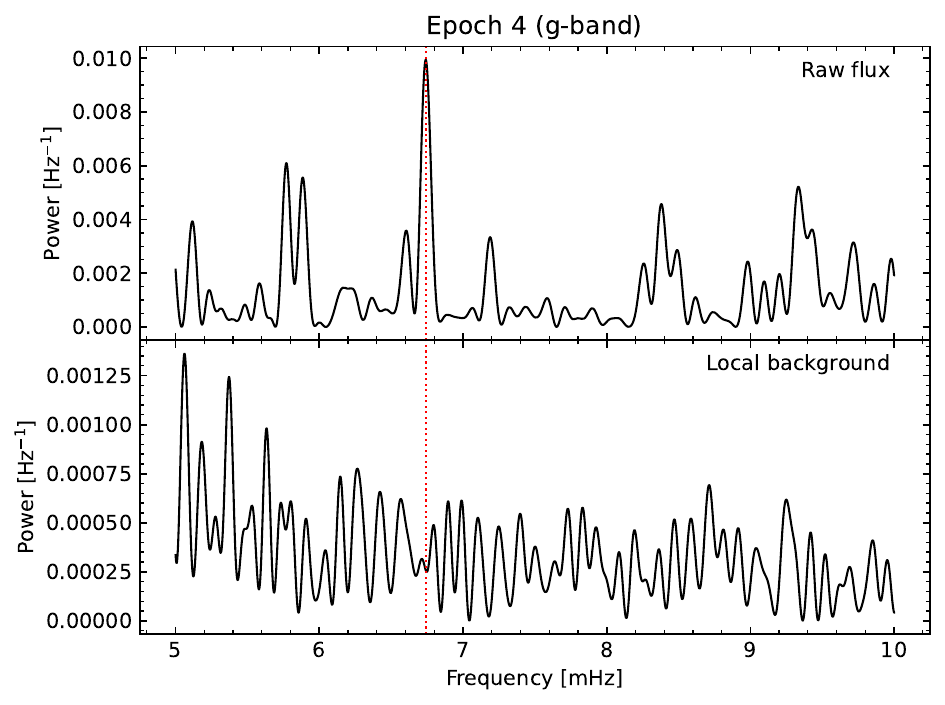}
    \caption{Top: LSP of \src's raw flux from Epoch 4 in the $g$-band; bottom: LSP of the corresponding local background.}
    \label{fig: epoch 4 local background periodogram}
\end{figure}

To test the veracity of the signal, we have computed LSPs of the other stars as well as the background itself, in case the signal was produced by artefacts from the electron noise in the detectors. We have not found any significant signal in any of these. In Figure \ref{fig: epoch 4 local background periodogram}, we present the LSP of the raw Epoch 4 $g$-band light curve for \src\ (top panel), along with the LSP of the corresponding local background (bottom panel). Figure \ref{fig: epoch 4 local background periodogram} shows a prominent peak in the periodogram of the raw flux at 6.7~mHz that is not seen in the periodogram of the local background. This rules out the 6.7~mHz signal being attributable to our reference stars, or a result of some instrumental/systematic effect, and instead suggests that the signal is intrinsic to \src.

\section{Discussion}

In Sec. \ref{sec: analysis}, we have unambiguously found evidence of a periodic signal with a period of $148.5 \pm 0.4$ s in epoch 4 during the dip in between the main outburst and the beginning of the rebrightenings/echo-outburst of the newly discovered WZ-Sge-type CV \src. Unfortunately, the weather conditions did not allow us to recover the signal from epoch 5, but the light curve from this night looks qualitatively similar. But, what is the origin of the 148.5 seconds signal observed in \src ? In this section, we will briefly explore the different physical mechanisms that can produce the observed signal. 

This signal is present in $g$-band with its amplitude decreasing toward longer wavelengths with a tentative detection in $r$-band. This suggests a blue spectral component generating the detected signal, similar to the one seen in IPs. In addition to this, the frequency of the observed signal is comparable with the dominant signal of a typical IP. This makes the \src\ a WZ Sge-type IP the most obvious interpretation, nonetheless we will consider other scenarios.

Dwarf nova oscillations (DNOs) have been proposed to explain some variable periodicity in outbursting CVs \citep[e.g.][]{Marsh1998, Woudt2002}, these are vertically extended regions of the disc or blobs being irradiated by the hot WD and boundary layer, which produce a signal corresponding to their Keplerian orbit. These signals are typically of the order of 10 seconds and evolve to slightly lower frequencies as the outburst evolves (up to 40s). We do not detect any significant signal during the main outburst, therefore we discard this scenario. 
The accretion-induced heating from the outburst makes non-radial pulsations of the WD also unlikely since the temperature of the WD should be too hot to be in the instability strip \citep[c.f. ][ but also see \citealt{Szkody_PulsatingAWDs:2021FrASS...8..184S}]{Clemens.ZZCeti.pulsatingWDs:1993BaltA...2..407C,Toloza.GWLib:2016MNRAS.459.3929T}, given its pre-outburst temperature \citep[][]{Killestein2025arXiv250111524K}. Finally, \citet{Veresvarska2024} proposed a precessing inner disc producing QPOs in a handful of CVs, however, these observed frequencies are typically one order of magnitude slower. Therefore we conclude that \src\ is likely a WZ Sge-IP, but, the stability of the signal would need to be tested once the system is back to quiescent levels. 

Magnetism has been invoked to explain several properties of WZ Sge stars (see Sect. \ref{sec: intro}). WZ Sge itself is probably the most remarkable example of this class. This source exhibits fast optical, UV and X-ray oscillations in quiescence \citep{Robinson1978, Patterson1980, Patterson1998, Skidmore1999} around 27.87s that have been associated with the spin period of the WD \citep{Patterson1998}. This signal is absent during outbursts owing to the combination of low magnetic field and/or low mass accretion rate from the donor, however, this interpretation has been challenged \citep[c.f. ][]{Knigge.WZSge2002ApJ...580L.151K}, leaving the origin of this signal an enduring enigma.
{\it CC Scl} and {\it ASASSN-18fk} are also proposed to harbour a slowly spinning WD \citep{Woudt2012MNRAS.427.1004W,Pavlenko:2019CoSka..49..204P,Paice_CCScl:2024MNRAS.531L..82P}. 
Another remarkable example of an intermediate polar exhibiting super-outbursts is {\it V455 And} \citep{V455And.Araujo-Betancor:2005A&A...430..629A, Bloemen.V455And:2013MNRAS.429.3433B}, this system exhibits a spin period of 67.6 seconds with the dominant signal in the power spectrum being twice the spin period; however, it does not exhibit echo outbursts like \src\ and {\it WZ Sge}. The timescale of the signal observed in \src\ falls in-between that of {\it CC Scl} and {\it V455 And}, and the absence of the signal during outburst could be attributed to the combination of low magnetic field and/or mass transfer rate 
similar to what has been proposed for {\it WZ Sge} \citep{Patterson1998}. We then propose the observed signal as being produced by the spin period (or its first harmonic).

The data from Epoch 4 was obtained 2-3 days after the decline from the main outburst; therefore, the disc and magnetic torques would have been far from equilibrium during this epoch, making any estimation of the magnetic field highly uncertain. However, we can set a lower limit on the magnetic field as the source’s magnetospheric radius must be larger than the WD radius. 
To do this, we need to estimate the mass accretion rate; for a canonical mass transfer rate of $\dot{\rm M}_{tr} \simeq 10^{-11}~\mathrm{M_\odot\,yr^{-1}}$ in WZ Sge stars \citep{Knigge+2011ApJS..194...28K}, typically expending $\gtrsim 30$ years in quiescence between outbursts, and with an outburst duration of $\simeq 30$~d, we estimate the average mass accretion rate during the outburst to be $<\dot{M}_{acc}>\simeq 7\times 10^{-9}~\mathrm{M_\odot\,yr^{-1}}$ in a conservative mass transfer scenario. Alternatively we can obtain the peak mass accretion rate via the peak absolute magnitude $\rm M_V$ vs orbital period ($P_{\rm orb}$) and its relation with $\dot{\rm M}_{acc}$ \citep[e.g. ][]{Warner_CVs_abs_mag:1987MNRAS.227...23W, Patterson:2011MNRAS.411.2695P}. 
For a source close to the period minimum (as \src\ is), we find the absolute magnitude at (super outburst) peak to be $\rm M_V \simeq 5$, consistent with the value derived by \citet{Killestein2025arXiv250111524K}. In turn, this value give us an estimation of the mass accretion rate of $\dot{\rm M}_{acc}\simeq 2\times 10^{-8}~\mathrm{M_\odot\,yr^{-1}}$ \citep[e.g.][]{Warner_CVs_abs_mag:1987MNRAS.227...23W}. We can therefore assume $\dot{\rm M}_{acc}\sim 10^{-8}~\mathrm{M_\odot\,yr^{-1}}$ a reasonable mass accretion rate during the main outburst of \src. 
During Epoch 4 the source was 2 orders of magnitude fainter, therefore, assuming the bulk observed luminosity during this epoch being dominated by the accretion disc and given that this scales with the mass accretion rate (${\rm L}_{acc}\propto\dot{\rm M}_{acc}$), we can adopt $\dot{\rm M}_{acc}\sim 10^{-10}~\mathrm{M_\odot\,yr^{-1}}$ during epoch 4 as a reasonable approximate reference value for our calculations (acknowledging that it excludes corrections or additional spectral components; e.g., the hot spot).
If the magnetosphere of the white dwarf truncates the disc at the co-rotation radius, i.e. $\rm r_M = \xi r_A = r_{CO}$ \citep[with $\rm \xi\simeq 0.5$ e.g. ][]{Ghosh:1979ApJ...234..296G, Long2005}, we can set a lower limit of the magnetic field, $\rm B$, at which $\rm r_M = R_{WD}$. For a typical $0.8~{\rm M_\odot}$ WD in a CV \citep{zorotovic2011,Pala2017MNRAS.466.2855P}. this would give us a lower limit of $\rm B\sim 2\times10^4~G$. This argument cannot be used for Epoch 3, since the disc is experiencing a rapid change which will keep it even further from a stationary configuration. The same is not true for Epoch 2, thus allowing us to set an upper limit in the magnetic field strength. Assuming the disc ``pushed'' magnetosphere to the surface of the WD during the outburst, we can constrain the magnetic field to be $B \lesssim 4\times10^4~G$.

In the hypothetical case of the disc and magnetic torques being in equilibrium during Epoch 4, we could estimate the magnetic field of the system, for $\rm P_{\omega} = 1$ and  $2 \times P_{spin}$ this would correspond to a magnetic field of $\rm 5 \times10^4~G$ and $\rm 10^5~G$ respectively. The latter of these values would require a $\dot{\rm M}_{acc}\simeq10^{-7}~\mathrm{M_\odot\,yr^{-1}}$ to push the magnetosphere down to the WD surface, this scenario not only goes against our constraints from above but it would make the system enter a super-Eddington wind regime \citep[c.f.][]{Ma_LEddWindsCVs:2013ApJ...778L..32M}.
We therefore suggest that $\rm P_\omega$ may be the fundamental frequency of the spin period, and the WD's magnetic would be of the order of $B\sim10^4~G$. However, quiescent observations are required to test this hypothesis.
If confirmed, \src\ would not only be another example of how magnetic fields can be dynamically important in highly evolved CVs but also would bridge the gap between intermediate- and fast-rotating AWDs.

\section*{Acknowledgements} \label{sec: discussion}
We are grateful to the referee for their helpful feedback. We are grateful to many amateur astronomers for making their data publicly available in {\it AAVSO}.
NCS and DLC acknowledges support from the Science and Technology Facilities Council (STFC) grant ST/X001121/1. ZAI, FV and DA acknowledges support STFC grants ST/V001000/1 and ST/X508767/1. IP acknowledges support from a Royal Society University Research Fellowship (URF\textbackslash R1\textbackslash 231496). AC acknowledges support from the Royal Society Newton International Fellowship grant NF170803. Based upon observations carried out at the Observatorio Astron\'omico Nacional on the Sierra San Pedro M\'artir (OAN-SPM), Baja California, M\'exico. We thank I. Zavala for maintaining the data acquisition software of OPTICAM.

\section*{Affiliations}
\noindent
{\it
$^{1}$Department of Physics, University of Warwick, Gibbet Hill Road, Coventry, CV4 7AL, UK\\
$^{2}$Department of Physics \& Astronomy. University of Southampton, Southampton SO17 1BJ, UK.\\
$^{3}$South African Astronomical Observatory, PO Box 9, Observatory, 7935, Cape Town, South Africa\\
$^{4}$Department of Astronomy, University of Cape Town, Private Bag X3, Rondebosch 7701, South Africa\\
$^{5}$Instituto de Astronom\'ia, Universidad Nacional Aut\'onoma de M\'exico, Carretera Tijuana-Ensenada Km. 107, Ensenada, B.C. 22860, M\'exico\\
$^{6}$SUPA School of Physics and Astronomy, University of St Andrews, North Haugh, St Andrews KY16 9SS, Scotland, UK\\
$^{7}$Centre for Extragalactic Astronomy, Department of Physics, Durham University, South Road, Durham, DH1 3LE\\
$^{10}$NAF -- Osservatorio Astronomico di Capodimonte, Salita Moiariello 16, I-80131 Naples, Italy\\
$^{8}$Advanced Research Center for Space Science and Technology, College of Science and Engineering, Kanazawa University, Kakuma, Kanazawa, Ishikawa 920-1192, Japan\\
}
\section*{Data Availability}
The data underlying this article is publicly available in {\em AAVSO} \url{https://www.aavso.org} and {\em ZTF} \url{https://www.ztf.caltech.edu/ztf-public-releases.html}. The remaining data will be shared on reasonable request to the corresponding author.




%
%





\bsp	
\label{lastpage}
\end{document}